\documentclass[man,12pt,noextraspace,floatsintext]{apa7}

\usepackage[american]{babel}
\usepackage[utf8]{inputenc}
\usepackage[T1]{fontenc}
\usepackage{amsfonts}
\usepackage{amsmath}
\usepackage{amssymb}
\usepackage{graphicx}
\usepackage{mathtools}
\usepackage{csquotes}
\usepackage[style=apa,sortcites=true,sorting=nyt,backend=biber]{biblatex}
\usepackage[normalem]{ulem}
\usepackage{xcolor}
\usepackage{hyperref}
\usepackage{multirow}
\usepackage{tablefootnote}
\usepackage{colortbl}

\DeclareLanguageMapping{american}{american-apa}

\newcommand{\RomanNumeralCaps}[1]{\MakeUppercase{\romannumeral #1}}

\title{baymedr: An R Package and Web Application for the Calculation of Bayes Factors for Superiority, Equivalence, and Non-Inferiority Designs}
\shorttitle{baymedr}

\leftheader{Linde, van Ravenzwaaij}

\authorsnames[1,1]{Maximilian~Linde,Don~van~Ravenzwaaij}
\authorsaffiliations{{Unit of Psychometrics and Statistics, Department of Psychology, Faculty of Behavioural and Social Sciences, University of Groningen, Groningen, The Netherlands}}

\abstract{Clinical trials often seek to determine the superiority, equivalence, or non-inferiority of an experimental condition (e.g., a new drug) compared to a control condition (e.g., a placebo or an already existing drug). The use of frequentist statistical methods to analyze data for these types of designs is ubiquitous even though they have several limitations. Bayesian inference remedies many of these shortcomings and allows for intuitive interpretations. In this article, we outline the frequentist conceptualization of superiority, equivalence, and non-inferiority designs and discuss its disadvantages. Subsequently, we explain how Bayes factors can be used to compare the relative plausibility of competing hypotheses. We present baymedr, an R package and web application, that provides user-friendly tools for the computation of Bayes factors for superiority, equivalence, and non-inferiority designs. Instructions on how to use baymedr are provided and an example illustrates how already existing results can be reanalyzed with baymedr.}


\authornote{
\addORCIDlink{Maximilian Linde}{0000-0001-8421-090X} \\
\addORCIDlink{Don van Ravenzwaaij}{0000-0002-5030-4091}\\
This research was supported by a Dutch scientific organization VIDI fellowship grant awarded to Don~van~Ravenzwaaij (016.Vidi.188.001). Correspondence concerning this article should be addressed to: Maximilian~Linde, University of Groningen, Department of Psychology,
Grote Kruisstraat 2/1, Heymans Building, room 217, 9712 TS Groningen, The Netherlands, Phone:~(+31)~50~363~2702, E-mail:~m.linde@rug.nl
}

\begin{document}

\maketitle

\section{Introduction}

Researchers generally agree that the clinical trial is the best method to determine and compare the effects of medications and treatments \parencite{Christensen2007, FriedmanFurbergDemets_2010}. Although clinical trials are often similar in design, different statistical procedures need to be employed depending on the nature of the research question. Commonly, clinical trials seek to determine the superiority, equivalence, or non-inferiority of an experimental condition (e.g., subjects receiving a new medication) compared to a control condition \parencite[e.g., subjects receiving a placebo or an already existing medication;][]{Lesaffre2008, PiaggioElbournePocock_2006}. For these goals, statistical inference is often conducted in the form of testing. 

Usually, the frequentist approach to statistical testing forms the framework in which data for these research designs are analyzed \parencite{ChavalariasWallachLi2016}. In particular, researchers often rely on null hypothesis significance testing (NHST), which quantifies evidence through a $p$-value. This $p$-value represents the probability of obtaining a test statistic (e.g., a $t$-value) at least as extreme as the one observed, assuming that the null hypothesis is true. In other words, the $p$-value is an indicator of the unusualness of the obtained test statistic under the null hypothesis, forming a ``proof by contradiction'' \parencite[][p.~123]{Christensen2005}. If the $p$-value is smaller than a predefined Type \RomanNumeralCaps 1 error rate ($\alpha$), typically set to $\alpha =.05$ \parencite[but see, e.g.,][]{BenjaminBergerJohannesson_2018, LakensAdolfiAlbers_2018}, rejection of the null hypothesis is warranted; otherwise the obtained data do not justify rejection of the null hypothesis.

The NHST approach to inference has been criticized due to certain limitations and erroneous interpretations of $p$-values \parencite[e.g.,][]{BergerSellke1987, Cohen1994, Dienes2011, GigerenzerKraussVitouch2004, Goodman1999a, Goodman2008, Loftus1996, Wagenmakers2007, WagenmakersMarsmanJamil_2018, WetzelsMatzkeLee_2011, Goodman1999b, VanravenzwaaijIoannidis2017, WassersteinLazar2016}, which we briefly describe below. As a result, some methodologists have argued that $p$-values should be mostly abandoned from scientific practice \parencite[e.g.,][]{WagenmakersMarsmanJamil_2018, BergerDelampady1987, Goodman2008, McshaneGalGelman_2019}.

An alternative to NHST is statistical testing within a Bayesian framework. Bayesian statistics is based on the idea that the credibilities of well-defined parameter values (e.g., effect size) or models (e.g., null and alternative hypotheses) are updated based on new observations \parencite{Kruschke2015}. With exploding computational power and the rise of Markov chain Monte Carlo methods \parencite[e.g.,][]{GilksRichardsonSpiegelhalter1995, VanravenzwaaijCasseyBrown2018} that are used to estimate probability distributions that cannot be determined analytically, applications of Bayesian inference have recently become tractable. Indeed, Bayesian methods are seeing more and more use in the biomedical field \parencite{Berry2006} and other disciplines \parencite{VandeschootWinterRyan_2017}.

Despite the fact that statistical inference is slowly changing from frequentist methods towards Bayesian methods, a majority of biomedical research still employs frequentist statistical techniques \parencite{ChavalariasWallachLi2016}. To some extent, this might be due to a biased statistical education in favor of frequentist inference. Moreover, researchers might perceive statistical inference through NHST and reporting of $p$-values as prescriptive and, hence, adhere to this convention \parencite{Winkler2001, Gigerenzer2004}. We believe that one of the most crucial factors is the unavailability of easy-to-use Bayesian tools and software, leaving Bayesian hypothesis testing largely to statistical experts. Fortunately, important advances have been made towards user-friendly interfaces for Bayesian analyses with the release of the BayesFactor software \parencite{MoreyRouder2018_0.9.12-4.2}, written in R \parencite{Rcoreteam2021}, and point-and-click software like JASP \parencite{Jaspteam2021_0.15} and Jamovi \parencite{Thejamoviproject2021_1.6}, the latter two of which are based to some extent on the BayesFactor software. However, these tools are mainly tailored towards research designs in the social sciences. Easy-to-use Bayesian tools and corresponding accessible software for the analysis of biomedical research designs specifically (e.g., superiority, equivalence, and non-inferiority) are still missing and, thus, urgently needed.

In this article, we provide a software package and a web application for conducting Bayesian hypothesis tests for superiority, equivalence, and non-inferiority designs. Although implementations for the superiority and equivalence test exist elsewhere, the implementation of the non-inferiority test is novel. Firstly, we outline the traditional frequentist approach to statistical testing for each of these designs. Secondly, we discuss the key disadvantages and potential pitfalls of this approach and motivate why Bayesian inferential techniques are better suited for these research designs. Thirdly, we explain the conceptual background of Bayes factors \parencite{Jeffreys1939, Jeffreys1948, Jeffreys1961, KassRaftery1995, Goodman1999b}. Fourthly, we provide and introduce baymedr \parencite{LindeVanravenzwaaij2021_0.1.1}, an open-source software written in R \parencite{Rcoreteam2021} that comes together with a web application (\href{https://maxlinde.shinyapps.io/baymedr/}{https://maxlinde.shinyapps.io/baymedr/}), for the computation of Bayes factors for common biomedical designs. We provide step-by-step instructions on how to use baymedr. Finally, we present a reanalysis of an existing empirical study to illustrate the most important features of the baymedr R package and the accompanying web application.

\section{Frequentist Inference for Superiority, Equivalence, and Non-Inferiority Designs}

The superiority, equivalence, and non-inferiority tests are concerned with research settings in which two conditions (e.g., control and experimental) are compared on some outcome measure \parencite{Lesaffre2008, Christensen2007}. For instance, researchers might want to investigate whether a new antidepressant medication is superior, equivalent, or non-inferior compared to a well-established antidepressant. For a continuous outcome variable, the between-group comparison is typically made with one or two $t$-tests. The three designs differ, however, in the precise specification of the $t$-tests (see Fig~\ref{fig:diagram}).

In the following, we will assume that higher scores on the outcome measure of interest represent a more favorable outcome (i.e., superiority or non-inferiority) than lower scores. For example, high scores are favorable when the measure of interest represents the number of social interactions in patients with social anxiety, whereas low scores are favorable when the outcome variable is the number of depressive symptoms in patients with major depressive disorder. We will also assume that the outcome variable is continuous and that the residuals within both conditions are Normal distributed in the population, sharing a common population variance. Throughout this article, the true population effect size ($\delta$) reflects the true standardized difference in the outcome between the experimental condition (i.e., $\text{e}$) and the control condition (i.e., $\text{c}$):
\begin{align}
    \delta = \dfrac{\mu _{\text{e}}- \mu _{\text{c}}}{\sigma} \text{.}
\end{align}

\subsection{The Superiority Design}

\begin{figure}[t!]
    \centering
    \includegraphics[width=0.95\textwidth]{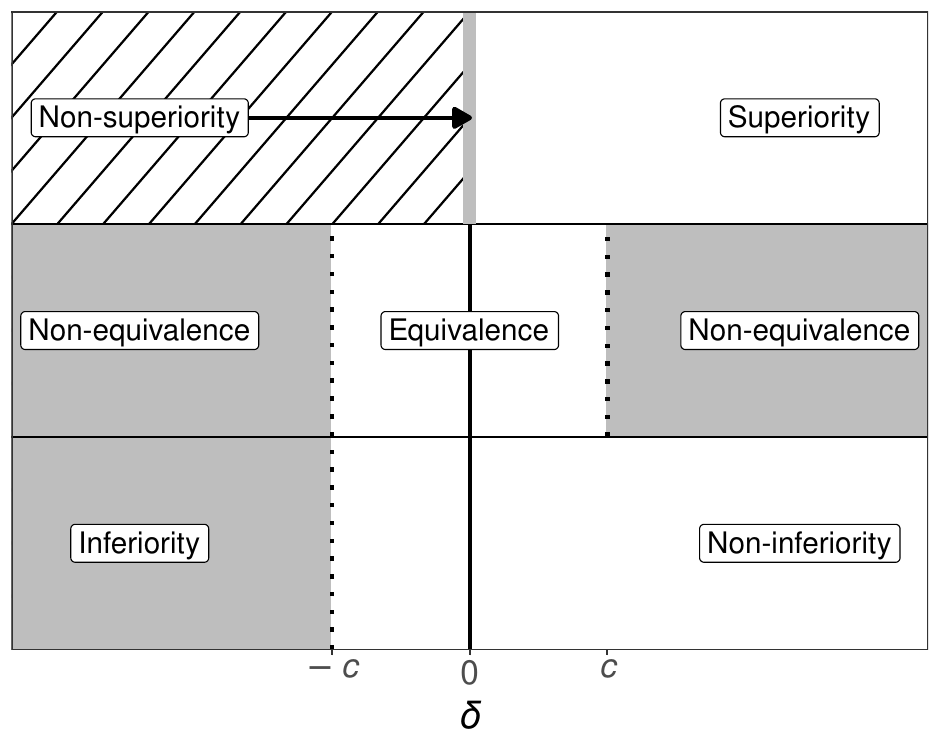}
    \caption{Schematic depiction of the superiority, equivalence, and non-inferiority designs. The $x$-axis represents the true population effect size ($\delta$), where $c$ is the standardized equivalence margin in case of the equivalence test and the standardized non-inferiority margin in case of the non-inferiority test. Gray regions mark the null hypotheses and white regions the alternative hypotheses. The region with the diagonal black lines is not used for the one-sided superiority design. Note that the diagram assumes that high values on the measure of interest represent superior or non-inferior values and that a one-sided test is used for the superiority design.}
    \label{fig:diagram}
\end{figure}

The superiority design tests whether the experimental condition is superior to the control condition (see the first row of Fig~\ref{fig:diagram}). Conceptually, the superiority design consists of a one-sided test due to its inherent directionality. The null hypothesis $\mathcal{H}_0$ states that the true population effect size is zero, whereas the alternative hypotheses $\mathcal{H}_1$ states that the true population effect size is larger than zero:
\begin{align}
    &\mathcal{H} _0 \mathpunct{:} ~ \delta =0&&\mathcal{H} _1 \mathpunct{:} ~ \delta >0 \text{.}
\end{align}
To test these hypotheses, a one-sided $t$-test is conducted.\footnote{Researchers often conduct a two-sided $t$-test and then confirm that the observed effect goes in the expected direction. We do not describe this approach because we have the opinion that a one-sided $t$-test should be conducted for the superiority test, whose name already implies a uni-directional alternative hypothesis.}

\subsection{The Equivalence Design}

The equivalence design tests whether the experimental and control conditions are practically equivalent (see the second row of Fig~\ref{fig:diagram}). There are multiple approaches to equivalence testing \parencite[see, e.g.,][]{Meyners2012}. A comprehensive treatment of all approaches is beyond the scope of this article. Here, we focus on one popular alternative: the two one-sided tests procedure \parencites[TOST;][]{HodgesLehmann1954, Westlake1976, Schuirmann1987}[see also][]{Senn2008, Meyners2012}. An equivalence interval must be defined, which can be based, for example, on the smallest effect size of interest \parencite{Lakens2017, LakensScheelIsager2018}. The specification of the equivalence interval is not a statistical question; thus, it should be set by experts in the respective fields \parencite{Meyners2012, Schuirmann1987} or comply with regulatory guidelines \parencite{Garrett2003}. Importantly, however, the equivalence interval should be determined independent of the obtained data. 

TOST involves conducting two one-sided $t$-tests, each one with its own null and alternative hypotheses. For the first test, the null hypothesis states that the true population effect size is smaller than the lower boundary of the equivalence interval, whereas the alternative hypothesis states that the true population effect size is larger than the lower boundary of the equivalence interval. For the second test, the null hypothesis states that the true population effect size is larger than the upper boundary of the equivalence interval, whereas the alternative hypothesis states that the true population effect size is smaller than the upper boundary of the equivalence interval. Assuming that the equivalence interval is symmetric around the null value, these hypotheses can be summarized as follows:
\begin{align}
    &\mathcal{H} _0 \mathpunct{:} \delta \leq -c ~ \text{OR} ~ \delta \geq c&&\mathcal{H} _1 \mathpunct{:} \delta > -c ~ \text{AND} ~ \delta < c \text{,}
\end{align}
where $c$ represents the margin of the standardized equivalence interval. Two $p$-values ($p_{-c}$ and $p_{c}$) result from the application of the TOST procedure. We reject the null hypothesis of non-equivalence and, thus, establish equivalence if $\max \left( p_{-c}, ~ p_{c} \right) < \alpha$ \parencite[cf.][]{Meyners2012, WalkerNowacki2011}. In other words, both tests need to reach statistical significance.

\subsection{The Non-Inferiority Design}

In some situations, researchers are interested in testing whether the experimental condition is non-inferior or not worse than the control condition by a certain amount. This is the goal of the non-inferiority design, which consists of a one-tailed test (see the third row of Fig~\ref{fig:diagram}). Realistic applications might include testing the effectiveness of a new medication that has fewer undesirable adverse effects \parencite{ChadwickVigabatrin1999}, is cheaper \parencite{KaulDiamond2006}, or is easier to administer than the current medication \parencite{VandewerfAdgeyArdissino_1999}. In these cases, we need to ponder the cost of a somewhat lower or equal effectiveness of the new treatment with the value of the just mentioned benefits \parencite{Hills2017}. The null hypothesis states that the true population effect size is equal to a predetermined threshold, whereas the alternative hypothesis states that the true population effect size is higher than this threshold:
\begin{align}
    &\mathcal{H} _0 \mathpunct{:} \delta =-c&&\mathcal{H} _1 \mathpunct{:} \delta >-c \text{,}
\end{align}
where $c$ represents the standardized non-inferiority margin. As with the equivalence interval, the non-inferiority margin should be defined independent of the obtained data.

\subsection{Limitations of Frequentist Inference}

Tests of superiority, equivalence, and non-inferiority have great value in biomedical research. It is the way researchers conduct their statistical analyses that, we argue, should be critically reconsidered. There are several disadvantages associated with the application of NHST to superiority, equivalence, and non-inferiority designs. Here, we limit our discussion to two disadvantages; for a more comprehensive exposition we refer the reader to other sources \parencite[e.g.,][]{Goodman1999a, Rennie1978, Internationalcommitteeofmedicaljournaleditors1997, WagenmakersMarsmanJamil_2018}.

First, researchers need to stick to a predetermined sampling plan \parencite{Rouder2014, SchonbrodtWagenmakersZehetleitner_2017, SchonbrodtWagenmakers2018}. That is, it is not legitimate to decide based on interim results to stop data collection (e.g., because the $p$-value is already smaller than $\alpha$) or to continue data collection beyond the predetermined sample size (e.g., because the $p$-value almost reaches statistical significance). In principle, researchers can correct for the fact that they inspected the data by reducing the required significance threshold through one of several techniques \parencite{RanganathanPrameshBuyse2016}. However, such correction methods are rarely applied. Especially in biomedical research, the possibility of optional stopping could reduce the waste of resources for expensive and time-consuming trials \parencite{ChalmersGlasziou2009}.

Second, with the traditional frequentist framework it is impossible to quantify evidence in favor of the null hypothesis \parencite{VanravenzwaaijMondenTendeiro_2019, Wagenmakers2007, WagenmakersMarsmanJamil_2018, Gallistel2009, RouderSpeckmanSun_2009}. Oftentimes, the $p$-value is erroneously interpreted as a posterior probability, in the sense that it represents the probability of the null hypothesis \parencite{BergerSellke1987, Gelman2013, Goodman2008, HallerKrauss2002}. However, a non-significant $p$-value does not only occur when the null hypothesis is in fact true but also when the alternative hypothesis is true, yet there was not enough power to detect an effect \parencite{Bakan1966, VanravenzwaaijMondenTendeiro_2019}. As \textcite[][p. 485]{AltmanBland1995} put it: ``Absence of evidence is not evidence of absence''. Still, a large proportion of biomedical studies falsely claim equivalence based on statistically non-significant $t$-tests \parencite{GreeneConcatoFeinstein2000}. Yet, estimating evidence in favor of the null hypothesis is essential for certain designs like the equivalence test \parencite{Blackwelder1982, VanravenzwaaijMondenTendeiro_2019, HoekstraMondenVanravenzwaaij_2018}.

The TOST procedure for equivalence testing provides a workaround for the problem that evidence for the null hypothesis cannot be quantified with frequentist techniques by defining an equivalence interval around $\delta =0$ and conducting two tests. Without this interval the TOST procedure would inevitably fail \parencite[see][for an explanation of why this is the case]{Meyners2012}. As we will see, the Bayesian equivalence test does not have this restriction; it allows for the specification of interval as well as point null hypotheses.

\section{Bayesian Tests for Superiority, Equivalence, and Non-Inferiority Designs}

The Bayesian statistical framework provides a logically sound method to update beliefs about parameters based on new data \parencite{Kruschke2015, Goodman1999b}. Bayesian inference can be divided into parameter estimation (e.g., estimating a population correlation) and model comparison (e.g., comparing the relative probabilities of the data under the null and alternative hypotheses) procedures \parencite[see, e.g.,][for an overview]{KruschkeLiddell2018b}. Here, we will focus on the latter approach, which is usually accomplished with Bayes factors \parencite{Jeffreys1939, Jeffreys1948, Jeffreys1961, KassRaftery1995, Goodman1999b}. In our exposition of Bayes factors in general and specifically for superiority, equivalence, and non-inferiority designs, we mostly refrain from complex equations and derivations. Formulas are only provided when we think that they help to communicate the ideas and concepts. We refer readers interested in the mathematics of Bayes factors to other sources \parencite[e.g.,][]{KassRaftery1995, Jeffreys1961, WagenmakersLodewyckxKuriyal_2010, OhaganForster2004, EtzVandekerckhove2018, RouderSpeckmanSun_2009}. The precise derivation of Bayes factors for superiority, equivalence, and non-inferiority designs in particular is treated elsewhere \parencites{VanravenzwaaijMondenTendeiro_2019}[see also][]{GronauLyWagenmakers2019}.

\subsection{The Bayes Factor}

Let us suppose that we have two hypotheses, $\mathcal{H} _0$ and $\mathcal{H} _1$, that we want to contrast. Without considering any data, we have initial beliefs about the probabilities of $\mathcal{H} _0$ and $\mathcal{H} _1$, which are given by the prior probabilities $p \left( \mathcal{H} _0 \right)$ and $p \left( \mathcal{H} _1 \right) =1-p \left( \mathcal{H} _0 \right)$. Now, we collect some data $D$. After having seen the data, we have new and refined beliefs about the probabilities that $\mathcal{H} _0$ and $\mathcal{H} _1$ are true, which are given by the posterior probabilities $p \left( \mathcal{H} _0 \mid D \right)$ and $p \left( \mathcal{H} _1 \mid D \right) =1-p \left( \mathcal{H} _0 \mid D \right)$. In other words, we update our prior beliefs about the probabilities of $\mathcal{H} _0$ and $\mathcal{H} _1$ by incorporating what the data dictates we should believe and arrive at our posterior beliefs. This relation is expressed in Bayes' rule:
\begin{equation} \label{eq_bayes_rule_hi}
    \underbrace{p \left( \mathcal{H} _i \mid D \right)}_{\text{Posterior}} = \dfrac{\overbrace{p \left( D \mid \mathcal{H} _i \right)}^{\text{Likelihood}} \overbrace{p \left( \mathcal{H} _i \right)}^{\text{Prior}}}{\underbrace{p \left( D \mid \mathcal{H} _0 \right) p \left( \mathcal{H} _0 \right) +p \left( D \mid \mathcal{H} _1 \right) p \left( \mathcal{H} _1 \right)}_{\text{Marginal Likelihood}}} \text{,}
\end{equation}
with $i= \{0, ~ 1 \}$, and where $p \left( \mathcal{H} _i \right)$ represents the prior probability of $\mathcal{H} _i$, $p \left( D \mid \mathcal{H} _i \right)$ denotes the likelihood of the data under $\mathcal{H} _i$, $p \left( D \mid \mathcal{H} _0 \right) p \left( \mathcal{H} _0 \right) +p \left( D \mid \mathcal{H} _1 \right) p \left( \mathcal{H} _1 \right)$ is the marginal likelihood \parencite[also called evidence; ][]{Kruschke2015}, and $p \left( \mathcal{H} _i \mid D \right)$ is the posterior probability of $\mathcal{H} _i$.

As we will see, the likelihood in Equation~\ref{eq_bayes_rule_hi} is actually a marginal likelihood because each model (i.e., $\mathcal{H}_0$ and $\mathcal{H}_1$) contains certain parameters that are integrated out. The denominator in Equation~\ref{eq_bayes_rule_hi} (labeled marginal likelihood) serves as a normalization constant, ensuring that the sum of the posterior probabilities is 1. Without this normalization constant the posterior is still proportional to the product of the likelihood and the prior. Therefore, for $\mathcal{H} _0$ and $\mathcal{H} _1$ we can also write:
\begin{equation}
    p \left( \mathcal{H} _i \mid D \right) \propto p \left( D \mid \mathcal{H} _i \right) p \left( \mathcal{H} _i \right) \text{,}
\end{equation}
where $\propto$ means ``is proportional to''.

Rather than using posterior probabilities for each hypothesis, let the ratio of the posterior probabilities for $\mathcal{H} _0$ and $\mathcal{H} _1$ be:
\begin{equation} \label{eq_post_odds}
    \underbrace{\dfrac{p \left( \mathcal{H} _0 \mid D \right)}{p \left( \mathcal{H} _1 \mid D \right)}}_{\text{Posterior odds}} = \underbrace{\dfrac{p \left( D \mid \mathcal{H} _0 \right)}{p \left( D \mid \mathcal{H} _1 \right)}}_{\text{Bayes factor,} ~ \text{BF}_{01}} \underbrace{\dfrac{p \left( \mathcal{H} _0 \right)}{p \left( \mathcal{H} _1 \right)}}_{\text{Prior odds}} \text{.}
\end{equation}
The quantity $p \left( \mathcal{H} _0 \mid D \right) /p \left( \mathcal{H} _1 \mid D \right)$ represents the posterior odds and the quantity $p \left( \mathcal{H} _0 \right) /p \left( \mathcal{H} _1 \right)$ is called the prior odds. To get the posterior odds, we have to multiply the prior odds with $p \left( D \mid \mathcal{H} _0 \right) /p \left( D \mid \mathcal{H} _1 \right)$, a quantity known as the Bayes factor \parencite{Jeffreys1939, Jeffreys1948, Jeffreys1961, KassRaftery1995, Goodman1999b}, which is a ratio of marginal likelihoods:
\begin{equation} \label{eq_marginal_likelihood}
    \text{BF}_{01}= \dfrac{\int _{\boldsymbol{\theta} _0} p \left( D \mid \boldsymbol{\theta} _0, \mathcal{H} _0 \right) p \left( \boldsymbol{\theta} _0 \mid \mathcal{H} _0 \right) d \boldsymbol{\theta} _0}{\int _{\boldsymbol{\theta} _1} p \left( D \mid \boldsymbol{\theta} _1, \mathcal{H} _1 \right) p \left( \boldsymbol{\theta} _1 \mid \mathcal{H} _1 \right) d \boldsymbol{\theta} _1} \text{,}
\end{equation}
where $\boldsymbol{\theta} _0$ and $\boldsymbol{\theta} _1$ are vectors of parameters under $\mathcal{H} _0$ and $\mathcal{H} _1$, respectively. In other words, the marginal likelihoods in the numerator and denominator of Equation~\ref{eq_marginal_likelihood} are weighted averages of the likelihoods, for which the weights are determined by the corresponding prior. In the case where one hypothesis has fixed values for the parameter vector $\boldsymbol{\theta} _i$ (e.g., a point null hypothesis), integration over the parameter space and the specification of a prior is not required. In that case, the marginal likelihood becomes a likelihood.

The Bayes factor is the amount by which we would update our prior odds to obtain the posterior odds, after taking into consideration the data. For example, if we had prior odds of 2 and the Bayes factor is 24, then the posterior odds would be 48. In the special case where the prior odds is 1, the Bayes factor is equal to the posterior odds. A major advantage of the Bayes factor is its ease of interpretation. For example, if the Bayes factor ($\text{BF}_{01}$, denoting the fact that $\mathcal{H} _0$ is in the numerator and $\mathcal{H} _1$ in the denominator) equals 10, the data are ten times more likely to have occurred under $\mathcal{H} _0$ compared to $\mathcal{H} _1$. With $\text{BF}_{01}=0.2$, we can say that the data are five times more likely under $\mathcal{H} _1$ compared to $\mathcal{H} _0$ because we can simply take the reciprocal of $\text{BF}_{01}$ (i.e., $\text{BF}_{10}=1/\text{BF}_{01}$). What constitutes enough evidence is subjective and certainly depends on the context. Nevertheless, rules of thumb for evidence thresholds have been proposed. For instance, \textcite{KassRaftery1995} labeled Bayes factors between 1 and 3 as ``not worth more than a bare mention'', Bayes factors between 3 and 20 as ``positive'', those between 20 and 150 as ``strong'', and anything above 150 as ``very strong'', with corresponding thresholds for the reciprocals of the Bayes factors. An alternative classification scheme was already proposed before, with thresholds at 3, 10, 30, and 100 and similar labels \parencites{Jeffreys1961}[see also][for updated labels]{LeeWagenmakers2013}.

Of course, we need to define $\mathcal{H} _0$ and $\mathcal{H} _1$. In other words, both models contain certain parameters for which we need to determine a prior distribution. Here, we will assume that the residuals of the two groups are Normal distributed in the population with a common population variance. The shape of a Normal distribution is fully determined with the location (mean; $\mu$) and the scale (variance; $\sigma ^2$) parameters. Thus, in principle, both models contain two parameters. Now, we make two important changes. 

Firstly, in the case where we have a point null hypothesis, $\mu$ under $\mathcal{H} _0$ is fixed at $\delta = 0$, leaving $\sigma ^2$ for $\mathcal{H} _0$ and $\mu$ and $\sigma ^2$ for $\mathcal{H} _1$. Parameter $\sigma ^2$ is a nuisance parameter because it is common to both models. Placing a Jeffreys prior (also called right Haar prior), $p \left( \sigma ^2 \right) \propto 1/ \sigma ^2$, on this nuisance parameter \parencite{Jeffreys1961, GonenJohnsonLu_2005, GronauLyWagenmakers2019} has several desirable properties that are explained elsewhere \parencite[e.g.,][]{BayarriBergerForte_2012, BergerPericchiVarshavsky1998}. 

Secondly, $\mu$ under $\mathcal{H} _1$ can be expressed in terms of a population effect size $\delta$ \parencite{RouderSpeckmanSun_2009, GonenJohnsonLu_2005}. This establishes a common and comparable scale across experiments and populations \parencite{RouderSpeckmanSun_2009}. The prior on $\delta$ could reflect certain hypotheses that we want to test. For instance, we could compare the null hypothesis ($\mathcal{H} _0 \mathpunct{:} \delta = 0$) to a two-sided alternative hypotheses ($\mathcal{H} _1 \mathpunct{:} \delta \neq 0$) or to one of two one-sided alternative hypotheses ($\mathcal{H} _1 \mathpunct{:} \delta < 0$ or $\mathcal{H} _1 \mathpunct{:} \delta > 0$). Alternatively, we could compare an interval hypothesis for the null hypothesis ($\mathcal{H} _0 \mathpunct{:} -c < \delta < c$) with a corresponding alternative hypothesis ($\mathcal{H} _1 \mathpunct{:} \delta < -c ~ \text{OR} ~ \delta > c$).\footnote{Note that the hypotheses represent exactly the opposite of the hypotheses in TOST (i.e., $\mathcal{H}_0$ of equivalence corresponds to $\mathcal{H}_1$ of equivalence in TOST, and vice versa). Evidence in favor of equivalence in TOST can only be obtained by rejecting two null hypotheses: $\mathcal{H}_0 \mathpunct{:} \delta < -c$ and $\mathcal{H}_0 \mathpunct{:} \delta > c$. For the Bayesian equivalence test we use the more intuitive null hypothesis of equivalence (i.e., $\mathcal{H}_0 \mathpunct{:} -c < \delta < c$).} The choice of the specific prior for $\delta$ is a delicate matter, which is discussed in the next section.

In the most general case, the Bayes factor (i.e., $\text{BF}_{01}$) can be calculated through division of the posterior odds by the prior odds (i.e., rearranging Equation~\ref{eq_post_odds}):
\begin{equation} \label{eq_bf01_gen}
    \text{BF}_{01}= \dfrac{\left( \dfrac{p \left( \mathcal{H} _0 \mid D \right)}{p \left( \mathcal{H} _1 \mid D \right)} \right)}{\left( \dfrac{p \left( \mathcal{H} _0 \right)}{p \left( \mathcal{H} _1 \right)} \right)} = \dfrac{\left( \dfrac{p \left( \mathcal{H} _0 \mid D \right)}{p \left( \mathcal{H} _0 \right)} \right)}{\left( \dfrac{p \left( \mathcal{H} _1 \mid D \right)}{p \left( \mathcal{H} _1 \right)} \right)} \text{;}
\end{equation}
accordingly, we can also calculate $\text{BF}_{10}$:
\begin{equation} \label{eq_bf10_gen}
    \text{BF}_{10}= \dfrac{\left( \dfrac{p \left( \mathcal{H} _1 \mid D \right)}{p \left( \mathcal{H} _0 \mid D \right)} \right)}{\left( \dfrac{p \left( \mathcal{H} _1 \right)}{p \left( \mathcal{H} _0 \right)} \right)} = \dfrac{\left( \dfrac{p \left( \mathcal{H} _1 \mid D \right)}{p \left( \mathcal{H} _1 \right)} \right)}{\left( \dfrac{p \left( \mathcal{H} _0 \mid D \right)}{p \left( \mathcal{H} _0 \right)} \right)} \text{.}
\end{equation}
Calculating Bayes factors this way often involves solving complex integrals \parencite[see, e.g., Equation~\ref{eq_marginal_likelihood}; also cf.][]{WagenmakersLodewyckxKuriyal_2010}. Fortunately, there is a computational shortcut for the specific but very common scenario where we have a point null hypothesis and a complementary interval alternative hypothesis. This shortcut, which is called the Savage-Dickey density ratio, takes the ratio of the density of the prior and posterior at the null value under the alternative hypothesis to calculate the Bayes factor; this is explained in more detail elsewhere \parencite{DickeyLientz1970, KassRaftery1995, WagenmakersLodewyckxKuriyal_2010, VanravenzwaaijEtz2021}.

\subsection{Default Priors}

Until this point in our exposition, we were quite vague about the form of the prior for $\delta$ under $\mathcal{H} _1$. In principle, the prior for $\delta$ within $\mathcal{H} _1$ can be defined as desired, conforming to the beliefs of the researcher. In fact, this is a fundamental part of Bayesian inference because various priors allow for the expression of a theory or prior beliefs \parencite{Vanpaemel2010, MoreyRomeijnRouder2016}. Most commonly, however, default or objective priors are employed that aim to increase the objectivity in specifying the prior or serve as a default when no specific prior information is available \parencite{Jeffreys1961, RouderSpeckmanSun_2009, ConsonniFouskakisLiseo_2018}. We employ objective priors in baymedr.

\begin{figure}[ht!]
    \centering
    \includegraphics[width=0.95\textwidth]{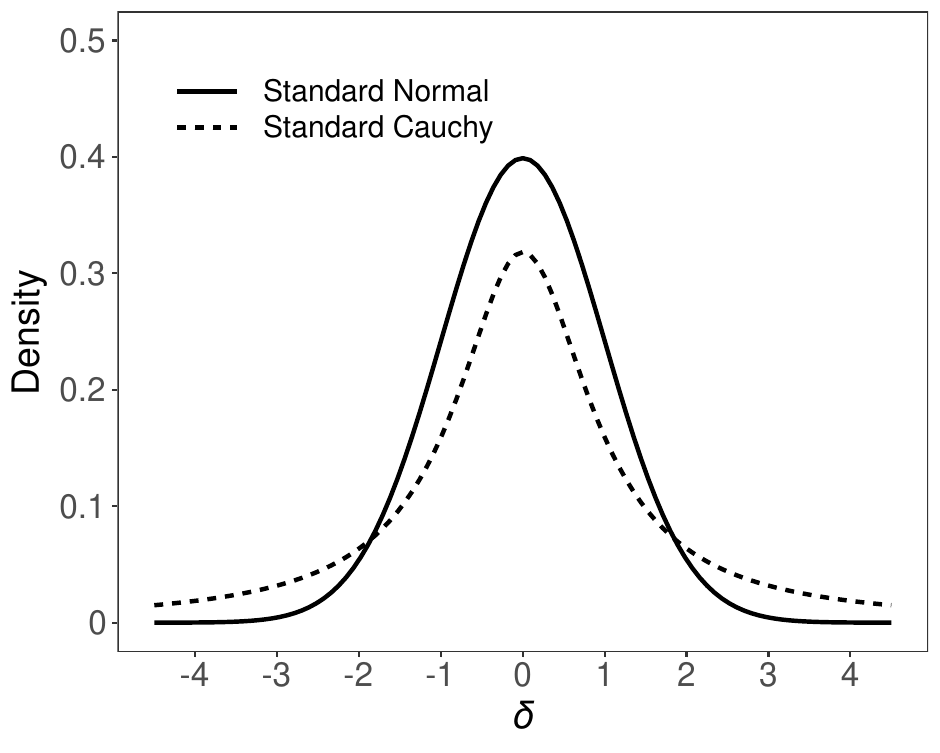}
    \caption{Comparison of the standard Normal probability density function (solid line) and the standard Cauchy probability density function (dashed line).}
    \label{fig:normal_cauchy}
\end{figure}

In the situation where we have a point null hypothesis and an alternative hypothesis that involves a range of values, \textcite{Jeffreys1961} proposed to use a Cauchy prior with a scale parameter of $r=1$ for $\delta$ under $\mathcal{H} _1$. This Cauchy distribution is equivalent to a Student's $t$ distribution with $1$ degree of freedom and resembles a standard Normal distribution, except that the Cauchy distribution has less mass at the center but instead heavier tails \parencite[see Fig~\ref{fig:normal_cauchy};][]{RouderSpeckmanSun_2009}. Mathematically, the Cauchy distribution corresponds to the combined specification of (1) a Normal prior with mean $\mu _ \delta$ and variance $\sigma ^2_ \delta$ on $\delta$; and (2) an inverse Chi-square distribution with $1$ degree of freedom on $\sigma ^2_ \delta$. Integrating out $\sigma ^2_ \delta$ yields the Cauchy distribution \parencite{LiangPauloMolina_2008, RouderSpeckmanSun_2009}. The scale parameter $r$ defines the width of the Cauchy distribution; that is, half of the mass lies between $-r$ and $r$.

Choosing a Cauchy prior with a location parameter of $0$ and a scale parameter of $r=1$ has the advantage that the resulting Bayes factor is $1$ in case of completely uninformative data. In turn, the Bayes factor approaches infinity (or $0$) for decisive data \parencite{Jeffreys1961, BayarriBergerForte_2012}. Still, by varying the Cauchy scale parameter, we can set a different emphasis on the prior credibility of a range of effect sizes. More recently, a Cauchy prior scale of $r=1/ \sqrt{2}$ is used as a default setting in the BayesFactor software \parencite{MoreyRouder2018_0.9.12-4.2}, the point-and-click software JASP \parencite{Jaspteam2021_0.15}, and Jamovi \parencite{Thejamoviproject2021_1.6}. We have adopted this value in baymedr as a default setting. Nevertheless, objective priors are often criticized \parencite[see, e.g.,][]{TendeiroKiers2019, KruschkeLiddell2018a}; researchers are encouraged to use more informed priors if relevant knowledge is available \parencite{RouderSpeckmanSun_2009, GronauLyWagenmakers2019}.

\section{Using baymedr}

With the baymedr software \parencite[BAYesian inference for MEDical designs in R;][]{LindeVanravenzwaaij2021_0.1.1}, written in R \parencite{Rcoreteam2021}, and the corresponding web application (accessible at \href{https://maxlinde.shinyapps.io/baymedr/}{https://maxlinde.shinyapps.io/baymedr/}) one can easily calculate Bayes factors for superiority, equivalence, and non-inferiority designs. The R package can be used by researchers who have only rudimentary knowledge of R; if that is not the case, researchers can use the web application, which does not require any knowledge of programming. In the following, we will demonstrate how Bayes factors for superiority, equivalence, and non-inferiority designs can be calculated with the baymedr R package; a thorough explanation of the web application is not necessary as it strongly overlaps with the R package. Subsequently, we will showcase (1) the baymedr R package and (2) the corresponding web application by reanalyzing data of an empirical study by \textcite{BasnerAschShea_2019}.

\subsection{The R Package}

\subsubsection{Install and Load baymedr}

To install the latest release of the baymedr R package from The Comprehensive R Archive Network (CRAN; \href{https://CRAN.R-project.org/package=baymedr}{https://CRAN.R-project.org/package=baymedr}), use the following command:
\begin{verbatim}
install.packages("baymedr")
\end{verbatim}
The most recent version of the R package can be obtained from GitHub (\href{https://github.com/maxlinde/baymedr}{https://github.com/maxlinde/baymedr}) with the help of the devtools package \parencite{WickhamHesterChang2019_2.2.0}:
\begin{verbatim}
devtools::install_github("maxlinde/baymedr")
\end{verbatim}
Once baymedr is installed, it needs to be loaded into memory, after which it is ready for usage:
\begin{verbatim}
library("baymedr")
\end{verbatim}

\subsubsection{Commonalities Across Designs}

For all three research designs, the user has three options for data input (function arguments that have ``x'' as a name or suffix refer to the control condition and those with ``y'' as a name or suffix to the experimental condition): (1) provide the raw data; the relevant arguments are \verb|x| and \verb|y|; (2) provide the sample sizes, sample means, and sample standard deviations; the relevant arguments are \verb|n_x| and \verb|n_y| for sample sizes, \verb|mean_x| and \verb|mean_y| for sample means, and \verb|sd_x| and \verb|sd_y| for sample standard deviations; (3) provide the sample sizes, sample means, and the confidence interval for the difference in group means; the relevant arguments are \verb|n_x| and \verb|n_y| for sample sizes, \verb|mean_x| and \verb|mean_y| for sample means, and \verb|ci_margin| for the confidence interval margin and \verb|ci_level| for the confidence level.

The Cauchy distribution is used as the prior for $\delta$ under the alternative hypothesis for all three tests. The user can set the width of the Cauchy prior with the \verb|prior_scale| argument, thus, allowing the specification of different ranges of plausible effect sizes. In all three cases, the Cauchy prior is centered on $\delta =0$. Further, baymedr uses a default Cauchy prior scale of $r=1/ \sqrt{2}$, complying with the standard settings of the BayesFactor software \parencite{MoreyRouder2018_0.9.12-4.2}, JASP \parencite{Jaspteam2021_0.15}, and Jamovi \parencite{Thejamoviproject2021_1.6}.

Once a superiority, equivalence, or non-inferiority test is conducted, an informative and accessible output message is printed in the console. For all three designs, this output states the type of test that was conducted and whether raw or summary data were used. Moreover, the corresponding null and alternative hypotheses are restated and the specified Cauchy prior scale is shown. In addition, the lower and upper bounds of the equivalence interval are presented in case an equivalence test was employed; similarly, the non-inferiority margin is printed when the non-inferiority design was chosen. Lastly, the resulting Bayes factor is shown. To avoid any confusion, it is declared in brackets whether the Bayes factor quantifies evidence towards the null (e.g., equivalence) or alternative (e.g., non-inferiority or superiority) hypothesis.

\subsubsection{Conducting Superiority, Equivalence, and Non-inferiority Tests}

The Bayesian superiority test is performed with the \verb|super_bf()| function. Depending on the research setting, low or high scores on the measure of interest represent ``superiority'', which is specified by the argument \verb|direction|. Since we seek to find evidence for the alternative hypothesis (superiority), the Bayes factor quantifies evidence for $\mathcal{H}_1$ relative to $\mathcal{H}_0$ (i.e., $\text{BF}_{10}$).

The Bayesian equivalence test is done with the \verb|equiv_bf()| function. The desired equivalence interval is specified with the \verb|interval| argument. Several options are possible: A symmetric equivalence interval around $\delta =0$ can be indicated by providing one value (e.g., \verb|interval = 0.2|) or by providing a vector with the negative and the positive values (e.g., \verb|interval = c(-0.2, 0.2)|). An asymmetric equivalence interval can be specified by providing a vector with the negative and the positive values (e.g., \verb|interval = c(-0.3, 0.2)|). The implementation of a point null hypothesis is achieved by using either \verb|interval = 0| or \verb|interval = c(0, 0)|, which also serves as the default specification. The argument \verb|interval_std| can be used to declare whether the equivalence interval was specified in standardized or unstandardized units. Since we seek to quantify evidence towards equivalence, we contrast the evidence for $\mathcal{H}_0$ relative to $\mathcal{H}_1$ (i.e., $\text{BF}_{01}$).

The Bayes factor for the non-inferiority design is calculated with the \verb|infer_bf()| function. The value for the non-inferiority margin can be specified with the \verb|ni_margin| argument. The argument \verb|ni_margin_std| can be used to declare whether the non-inferiority margin was given in standardized or unstandardized units. Lastly, depending on whether high or low values on the measure of interest represent ``non-inferiority'', one of the options ``high'' or ``low'' should be set for the argument \verb|direction|. We wish to determine the evidence in favor of $\mathcal{H}_1$; therefore, the evidence is expressed for $\mathcal{H}_1$ relative to $\mathcal{H}_0$ (i.e., $\text{BF}_{10}$).

\subsection{Demonstration of baymedr}

To illustrate how the R package and the web application can be used, we provide one example of an empirical study that employed non-inferiority tests to investigate differences in the amount of sleep, sleepiness, and alertness among medical trainees following either standard or flexible duty-hour programs \parencite{BasnerAschShea_2019}. The authors list several disadvantages of restricted duty-hour programs, such as: (1) ``[t]ransitions [as a result of restricted duty hours] into and out of night shifts can result in fatigue from shift-work-related sleep loss and circadian misalignment''; (2) ``[p]reventing interns from participating in extended shifts may reduce educational opportunities''; (3) ``increase[d] handoffs''; (4) ``reduce[d] continuity of care''; and (5) ``[r]estricting duty hours may increase the necessity of cross-coverage, contributing to work compression for both interns and more senior residents'' \parencite[][p. 916]{BasnerAschShea_2019}. As outlined above, the calculation of Bayes factors for equivalence and superiority tests is done quite similarly to the non-inferiority test, so we do not provide specific examples for those tests. For the purpose of this demonstration, we will only consider the outcome variable sleepiness. Participants were monitored over a period of $14$ days and were asked to indicate each morning how sleepy they were by completing the Karolinska sleepiness scale \parencite{AkerstedtGillberg1990}, a $9$-point Likert scale ranging from $1$ (extremely alert) to $9$ (extremely sleepy, fighting sleep). The dependent variable consisted of the average sleepiness score over the whole observation period of $14$ days. The research question was whether the flexible duty-hour program was non-inferior to the standard program in terms of sleepiness.

The null hypothesis was that medical trainees in the flexible program are sleepier by more than a non-inferiority margin than trainees in the standard program. Conversely, the alternative hypothesis was that trainees in the flexible program are not sleepier by more than a non-inferiority margin than trainees in the standard program. The non-inferiority margin was defined as $1$ point on the $9$-point Likert scale. All relevant summary statistics can be obtained or calculated from Table~$1$ and the Results section of \textcite{BasnerAschShea_2019}. Table~$1$ indicates that the flexible program had a mean of $M_e=4.8$ and the standard program had a mean of $M_c=4.7$. From the Results section we can extract that sample sizes were $n_e=205$ and $n_c=193$ in the flexible and standard programs, respectively. Further, the margin of the $95 \%$~CI of the difference between the two conditions was $0.31-0.12=0.19$. Finally, lower scores on the sleepiness scale constitute favorable (non-inferior) outcomes.

\subsubsection{The R Package}

Using this information, we can use the baymedr R package to calculate the Bayes factor as follows:
\begin{verbatim}
infer_bf(n_x = 193, n_y = 205,
         mean_x = 4.7, mean_y = 4.8,
         ci_margin = 0.19, ci_level = 0.95,
         ni_margin = 1, ni_margin_std = FALSE,
         prior_scale = 1 / sqrt(2),
         direction = "low")
\end{verbatim}
Note that we decided to use a Cauchy prior scale of $r=1/ \sqrt{2}$ for this reanalysis. Since our Cauchy prior scale of choice represents the default value in baymedr, it would not have been necessary to provide this argument; however, for purposes of illustration, we mentioned it explicitly in the function call.

The output provides a user-friendly summary of the analysis:
\begin{verbatim}
******************************
Non-inferiority analysis
------------------------
Data:                         summary data
H+ (inferiority):             mu_y - mu_x > ni_margin
H- (non-inferiority):         mu_y - mu_x < ni_margin
Non-inferiority margin:       1.04 (standardised)
                              1.00 (unstandardised)
Cauchy prior scale:           0.707

    BF-+ (non-inferiority) = 8.56e+10
******************************
\end{verbatim}
This large Bayes factor supports the conclusion from \textcite{BasnerAschShea_2019} that medical trainees in the flexible duty-hour program are non-inferior in terms of sleepiness compared to medical trainees in the standard program ($p<.001$). In other words, the data are $8.56 \times {10}^{10}$ more likely to have occurred under $\mathcal{H}_1$ than $\mathcal{H}_0$.

\subsection{The Web Application}

\begin{figure}[t!]
    \centering
    \includegraphics[width=0.95\textwidth]{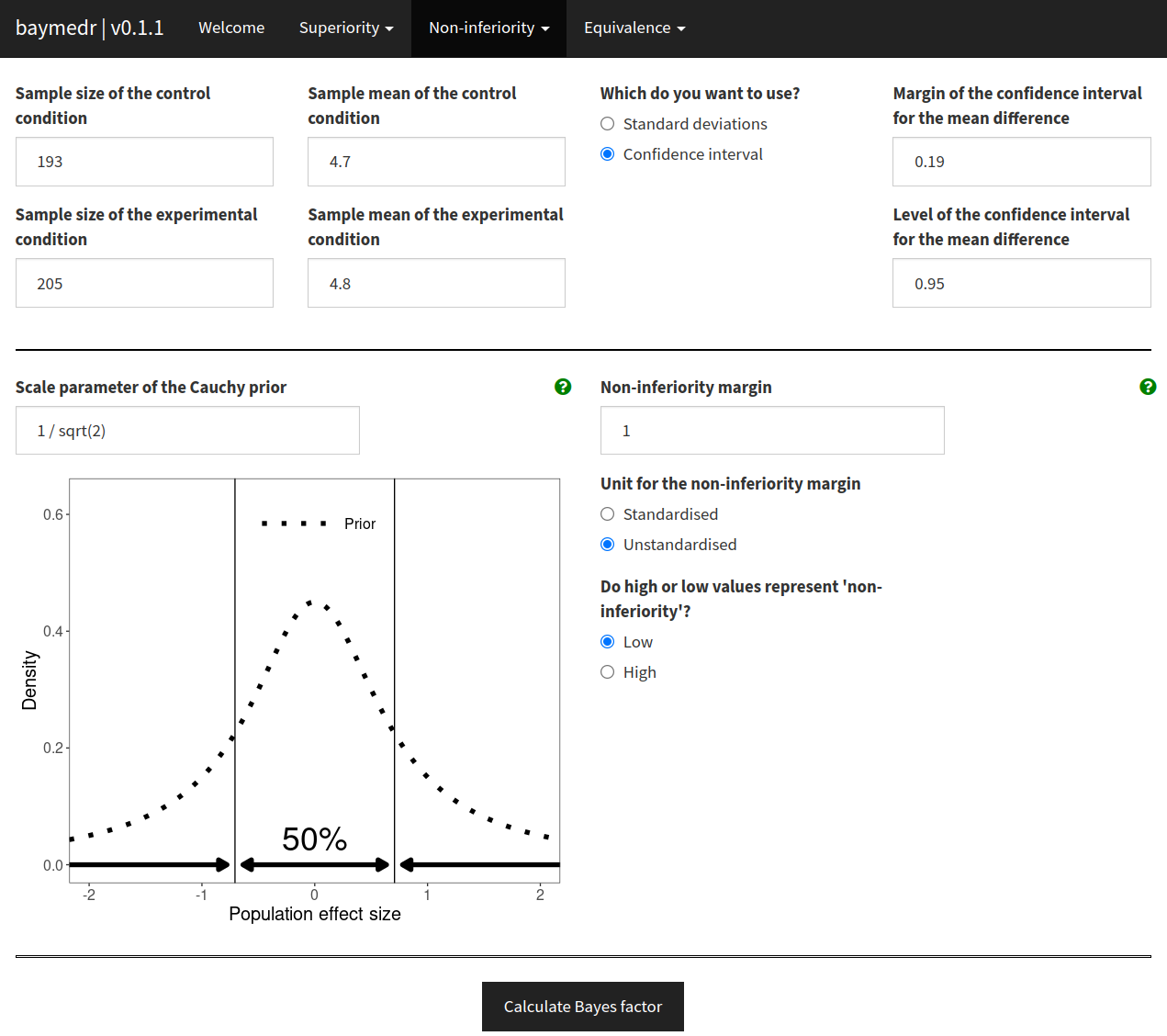}
    \caption{Shown is part of the baymedr web application demonstrating how summary statistics can be inserted and further parameters specified for a Bayesian non-inferiority test. In this specific case, the summary statistics correspond to the ones obtained from \textcite{BasnerAschShea_2019}. See text for details.}
    \label{fig:shiny_input}
\end{figure}

Similarly, we can use the web application to calculate the Bayes factor. For this, the web application should first be opened in a web browser (\href{https://maxlinde.shinyapps.io/baymedr/}{https://maxlinde.shinyapps.io/baymedr/}). The just-opened welcome page offers a brief description of the three research designs and Bayes factors and lists several further useful resources for the interested user. Since we want to conduct a non-inferiority test with summary data, we click on ``Non-inferiority'' and then ``Summary data'' on the navigation bar at the top (see Fig~\ref{fig:shiny_input}). The summary statistics for the example reanalysis of \textcite{BasnerAschShea_2019} can be inserted in the corresponding fields, as shown in Fig~\ref{fig:shiny_input}. For some fields a small green question mark is shown, which provides more details and help when the user clicks on them. Furthermore, the scale of the prior distribution can be specified, which by default is set to \verb|1 / sqrt(2)|. A small dynamic plot accompanies the field for the Cauchy prior scale. That is, once the prior scale is changed, the plot updates automatically, so that users obtain an impression of what the distribution looks like and what effect sizes are included. Once the ``Calculate Bayes factor'' button is clicked, the output is displayed.

\begin{figure}[t!]
    \centering
    \includegraphics[width=0.95\textwidth]{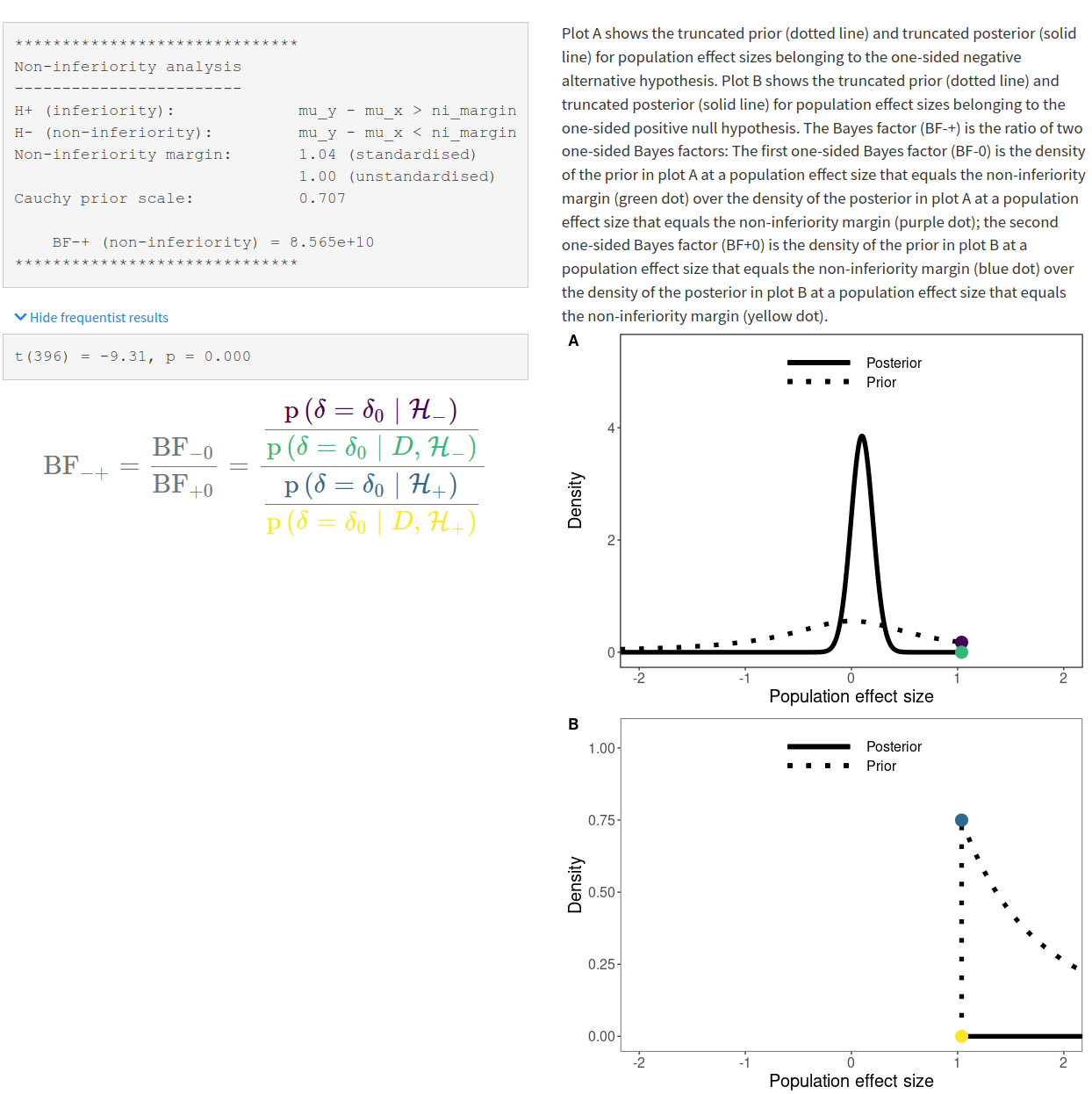}
    \caption{Shown is part of the baymedr web application showing the results of a Bayesian non-inferiority test. In this specific case, the results correspond to a reanalysis using summary statistics obtained from \textcite{BasnerAschShea_2019}. See text for details.}
    \label{fig:shiny_output}
\end{figure}

Fig~\ref{fig:shiny_output} shows the output of the calculations. The top of the left column displays the same output that is given with the R package. Further, upon clicking on ``Show frequentist results'', the results of the frequentist non-inferiority test are shown and clicking on ``Hide frequentist results'' in turn hides those results. Below that output is the formula for the Bayes factor, with different elements printed in colors that correspond to dots in matching colors in the plots on the right column of the results output. The upper plot shows the prior and posterior for contrasting $\mathcal{H}_0 \mathpunct{:} \delta = c$ with $\mathcal{H}_1 \mathpunct{:} \delta < c$. The two distributions are truncated, meaning that they are cut off at $\delta = c$. Similarly, the lower plot shows the truncated prior and posterior for contrasting $\mathcal{H}_0 \mathpunct{:} \delta = c$ with $\mathcal{H}_1 \mathpunct{:} \delta > c$. Through a heuristic called the Savage-Dickey density ratio \parencite{DickeyLientz1970, KassRaftery1995, WagenmakersLodewyckxKuriyal_2010, VanravenzwaaijEtz2021}, the ratio of the heights of the colored dots gives us the Bayes factor (see the colored expressions in the formula on the right side of the results output). The text above the two plots explains the plots as well.

\section{Discussion}

Tests of superiority, equivalence, and non-inferiority are important means to compare the effectiveness of medications and treatments in biomedical research. Despite several limitations, researchers overwhelmingly rely on traditional frequentist inference to analyze the corresponding data for these research designs \parencite{ChavalariasWallachLi2016}. We believe that Bayes factors \parencite{Jeffreys1939, Jeffreys1948, Jeffreys1961, KassRaftery1995, Goodman1999b} are an attractive alternative to NHST and $p$-values because they allow researchers to quantify evidence in favor of the null hypothesis \parencite{VanravenzwaaijMondenTendeiro_2019, WagenmakersMarsmanJamil_2018, Wagenmakers2007, Gallistel2009} and permit sequential testing and optional stopping \parencite{Rouder2014, SchonbrodtWagenmakersZehetleitner_2017, SchonbrodtWagenmakers2018}. In fact, we believe that the possibility for optional stopping and sequential testing has the potential to largely reduce the waste of scarce resources. This is especially important in the field of biomedicine, where clinical trials might be expensive or even harmful for participants.

Our baymedr R package and web application \parencite{LindeVanravenzwaaij2021_0.1.1} enable researchers to conduct Bayesian superiority, equivalence, and non-inferiority tests. baymedr is characterized by a user-friendly implementation, making it convenient for researchers who are not statistical experts. Furthermore, using baymedr, it is possible to calculate Bayes factors based on raw data and summary statistics, allowing for the reanalysis of published studies, for which the full data set is not available.

\section{Competing interests}

The authors declare no competing interests.

\printbibliography

\end{document}